\documentstyle[epsfig,twocolumn]{mn}

\title[A cosmic ray cocoon along the X-ray jet of M87?]{A cosmic ray cocoon along M87's jet?}
\author[M.G. Dainotti et al.]{M.G. Dainotti$^{1,3}$, M. Ostrowski $^{1}$, D. Harris $^{2}$, A. Siemiginowska $^{2}$, H. Siejkowski $^1$\\
$^1$ Obserwatorium Astronomiczne, Uniwersytet Jagiello{\'n}ski, ul. Orla 171, 30-244 Krak{\'o}w, Poland \\ $^2$ Smithsonian Astrophysical Observatory, 60 Garden Street, Cambridge, MA 02138, USA}

\date{Accepted xxx, Received yyy, in original form zzz}
\begin{document}
\maketitle
\begin{abstract}

Relativistic jets propagating through an ambient medium must produce
some observational effects along their side boundaries because of
interactions across the large velocity gradient. One possible effect
of such an interaction would be a sheared magnetic field structure at
the jet boundaries, leading to a characteristic radio polarization
pattern. As proposed by Ostrowski, another effect can come from
the generation of a high energy cosmic ray component at the boundary,
producing dynamic effects on the medium surrounding the jet and
forming a cocoon dominated by cosmic rays with a decreased thermal gas
emissivity. We selected this process for our first attempt to look for
the effects of this type of interaction. We analyzed the Chandra X-ray
data for the radio galaxy M87 in order to verify if the expected
regions of diminished emissivity may be present near the spectacular
X-ray jet in this source. The detailed analysis of the data, merged
from 42 separate observations, shows signatures of lower emissivity
surrounding the jet. In particular we detect an intensity dip along
the part of the jet, which would be approximately 150 pc x 2 kpc in
size, if situated along the jet which is inclined toward us. Due to a
highly non-uniform X-ray background in the central region we are not
able to claim the discovery of a cosmic ray cocoon around the M87
jet: we only have demonstrated that the data show morphological
structures which could be accounted for if a cosmic ray cocoon exists.
\end{abstract}
\section{Introduction}

M87 (NGC 4486) is a dominant radio galaxy in the center of the Virgo
cluster, at a distance of D = 16 Mpc \cite{Tonry1991} so that
$1^{\prime\prime}=77$ pc. It hosts a $3.2\times10^9$ solar masses
supermassive black hole \cite{Harms1994,Ford1994,Macchetto1997} and a
relativistic jet
\cite{Spark1996,Perlman2001,Marshall2002,Harris2003}. The first
Chandra observations of M87 that specifically targeted the jet were
those by Wilson in 2000 July \cite{Wilson2002,Perlman2005}. On larger
scales M87 has been the subject of detailed radio observations showing
remarkable structures on scales up to 40 kpc \cite{Owen2000,Hines
  1989}. In soft X-rays, M87 is the second brightest extragalactic
source (after the Perseus cluster) and the emission is dominated by
the thermal radiation from its $\sim 2$ keV gaseous atmosphere c.f.,
\cite{Molendi2002,Gorenstein1997,Fabricant1983,Bohringer2001,Matsushita2002,Belsole2001}.

Here we investigate the surface brightness profiles across the M87 jet
in the X-ray band $(0.3-7.0 \, {\rm keV})$, observed by Chandra
to search for possible effects of a relativistic jet on the
ambient medium of the host galaxy caused by high-energy cosmic rays (CRs)
accelerated at the jet side boundary, see Ostrowski (2000). In extreme
cases, such accelerated cosmic rays are expected to be able to push
the ambient medium off the jet, creating a `cosmic ray dominated
cocoon', separating the jet from the surrounding medium. The mechanism
becomes efficient provided the velocity difference across
the jet boundary is relativistic and a sufficient amount of
turbulence is generated in the medium; see Ostrowski
\shortcite{Ostrowski1990,Ostrowski1998,Ostrowski2000}.

Dynamic effects caused by CRs depend on the ratio of the CR pressure,
$P_{\mathrm{cr}}$, to the ambient medium pressure, $P_{\mathrm
  {ext}}$, at the jet boundary. If, at small CR energies, the
acceleration time scale is longer than the jet dynamic time or the
particle escape is efficient, the resulting energetic particle
population cannot reach a sufficiently high energy density to produce
dynamic effects. In this case CRs act only as a viscous agent near the
boundary, decreasing slightly the gas and magnetic field concentration
in this region, see also in Arav \& Begelman
\shortcite{AravBegelman}. In such cases we call the cylindrically
distributed cosmic ray population a `{\it weak} cosmic ray
cocoon'. Then, if the accelerated particles are electrons or can
transfer energy to electrons, a specific cosmic ray electron
population is formed along the jet leading to a characteristic
synchrotron component with slowly varying spectral index and break
frequency. The energy density of such radiating electrons is expected
to have its maximum in a cylindrical layer at the jet
boundary\footnote{Or at relativistic velocity jump interfaces
  between jet components, as proposed previously
  \cite{Meliani2009}}. However, if the energy losses are not
significant at low energies and the acceleration process is fast
enough, the cosmic ray pressure could reach values comparable to the
medium pressure, leading to a substantial modification of the jet
boundary layer. There are various possibilities arising in such cases
(`{\it dynamic cosmic ray cocoons}' \cite{Ostrowski2000})
depending on the conditions near the jet and on the effects of a CR
backreaction on the efficiency of the acceleration process. The cosmic ray
pressure may stabilize at an intermediate pressure $P_{\mathrm{cr}}
<P_{\mathrm {ext}}$, or grow to form the dynamic cosmic ray cocoon, as
one can see in Fig. 3 in Ostrowski (2000).

Let us consider a possible scenario of dynamic interaction of high
energy cosmic rays with the jet and the ambient medium. The particles
are injected and accelerated at the jet boundary. Growing numbers of
such particles result in forming the cosmic ray pressure gradient
outside the jet pushing the ambient medium aside. Additionally, an
analogous gradient may be formed directly into the jet, helping to
keep it collimated. The resulting rarefied or partially empty portion
of the ambient medium cylindrical volume along the jet will decrease
the acceleration efficiency. Thus the cosmic ray energy density may
build up only to a value comparable to the ambient medium pressure,
when it is then able to push the ambient medium away, because the diffusive
escape of charged particles from the cosmic ray cocoon is not expected
to be efficient (contrary to photons considered by Arav \& Begelman
\shortcite{AravBegelman}). In some cases the evacuated volumes could
be quite large, reaching values comparable to the jet's radius, or
even larger, expanding into the hot ambient gas. The accumulated
cosmic rays can be removed by advection in the form of CR
bubbles or CR dominated winds outside the active
nucleus, moving into regions of more tenuous medium, or they can simply
diffuse outside the jet at larger distances from the central
source. In all of these cases, one may expect a decrease of the
thermal emissivity near the jet -- in the CR dominated cocoon
-- leaving an observational signature of the CR related process. In
the present paper we attempt to verify such a model using X-ray
observations of the jet in the galaxy M87, searching for a lower
emissivity volume along the jet.  In Section 2 we
present the data analysis, in Section 3 the evidence and explanation
of the soft X-ray structures in the M87 central region, in Section 4
the physical constraints for the cosmic ray cocoon and in the final
remarks we discuss the existence of an X-ray morphological
structure along the M87 jet, which could be an indication of a cosmic
ray cocoon.

\section{Data Analysis}

We have selected 42 observations of M87 obtained by the Chandra AXAF
CCD Imaging Spectrometer (ACIS).  Amongst these are the longest
observations but we have also included the majority of the 5ks ACIS-S
monitoring observations with 0.4s frame time.  Since we focused our
analysis on a jet segment far enough removed from HST-1, we only
avoided using 5ks observations during the time HST-1 was extremely
bright. The total observing time is $\approx 634$
ks. In Table \ref{data}, information about the analyzed data is
presented.

\begin{figure}
\centering
\includegraphics[width=1.05\hsize,angle=0,clip]{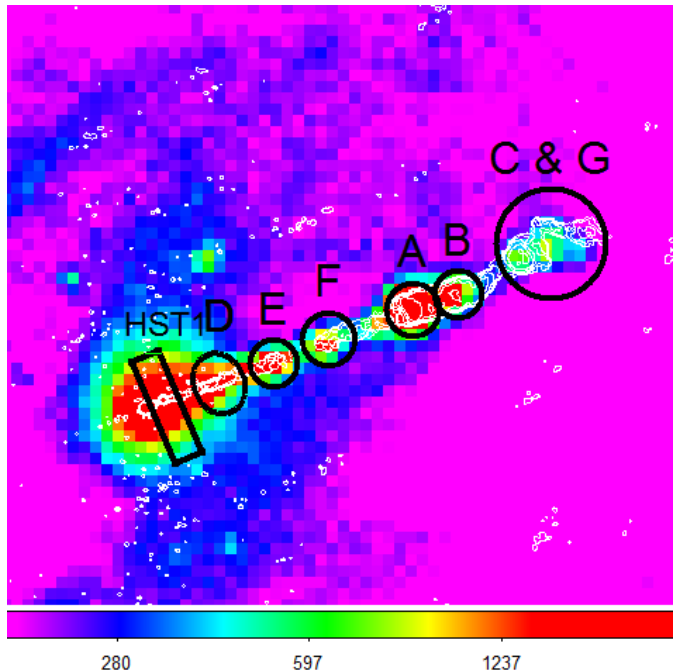}
\caption{The merged image without any smoothing or binning.  The
  energy band is $0.3-2.0$ keV and VLA radio contours are
  superimposed. In the picture, between knots E and A, one can note a
  dip in X-ray intensity on the north side of the jet. A circle around
  knot F shows a region used for verification of the data merging
  procedure. Distances of the knots from the central black hole:
  HST-1: $0.8''-1.2''$, D: $2.7''-4''$, E: $5.7''-6.2''$,
  F:$8.1''-8.8''$, I: $10.5''-11.5''$, A: $12.2''-12.5''$, B:
  $14.1''-14.5''$, C: $17.5''-19''$. Every pixel in any image is
  $0.492"$ and $1"=77$ pc, the projected length of the jet from the
  core to knot C is $\approx 1.5$ kpc.}
\label{fig2}
\end{figure}

We started our analysis with the longest single observation obsid
5827. Our results showed a region near the jet with lower surface
brightness between knots E and A, Fig. \ref{fig2}. Motivated by this
first result we merged all the available data to increase
considerably the number of counts in order to perform a detailed analysis of
the emission surrounding the jet.

\begin{table}
\caption{A list of observations and their exposure times used in the present analysis, with a separation of the chip 7 data (top) and the chip 3 data (bottom).}
\begin{center}
\begin{tabular}{|c|c|c|}
\hline
$Obs ID$ & $ Exposure$ & $Date$\\
 & [ks]& \\
\hline
10282& 4.70& 2008-11-17 \\
10283& 4.70& 2009-01-07 \\
10284& 4.70& 2009-02-20\\
10285& 4.66& 2009-04-01\\
10286& 4.68& 2009-05-13  \\
10287& 4.70& 2009-06-22\\
1808 & 14.7 & 2000-07-30\\
3084 &5.13 & 2002-02-12\\
3085 &5.39 & 2002-01-16 \\
3086 & 5.09 & 2002-03-30\\
3087  &	5.48 &2002-06-08\\
3088  &	5.19 &2002-07-24 \\
3975 & 5.83 & 2002-11-17\\
3976 & 5.28& 2002-12-29\\
3977& 5.82& 2003-02-04 \\
3978& 5.35& 2003-03-09 \\
3979& 4.95& 2003-04-14 \\
3980& 5.28& 2003-05-18 \\
7350& 5.14& 2007-02-13\\
7351& 5.16& 2007-03-24\\
7352& 5.06& 2007-05-15\\
7353& 5.01& 2007-06-25\\
7354& 5.19& 2007-07-31\\ 
8510& 4.70& 2007-02-15\\
8511&4.70& 2007-02-18\\
8512&4.70& 2007-02-21\\
8513&4.70& 2007-02-24\\
8514&4.47& 2007-03-12\\
8515&4.69& 2007-03-14\\
8516&4.67& 2007-03-19\\
8517&4.67& 2007-03-22\\
8575& 5.16&  2007-11-25  \\ 
8576& 5.17&  2008-01-04  \\ 
8577& 5.14&  2008-02-16  \\ 
8578& 5.19&  2008-04-01  \\ 
8579& 5.19& 2008-05-15  \\ 
8580& 5.19&  2008-06-24  \\ 
8581& 5.13& 2008-08-07  \\ 
2707&99.93& 2002-07-06 \\
\hline
7212&66.11& 2005-11-14 \\
5826&128.44&2005-03-03 \\
5827&158.27& 2005-05-05 \\
\hline
\end{tabular}
\label{data}
\end{center}
\end{table}
  
\subsection{Merging procedure}

Before starting the data merging procedure in all observations, we
corrected the nominal RA and Dec in the fits header in order to ensure
that the celestial location of the observed nuclear X-ray emission
coincided with the radio position of the nucleus. This procedure
didn't change the physical coordinates of events. Then, we applied the
simple merging procedure ``merge all'' from \textsc{ciao} (Chandra Interactive
Analysis of Observations)\footnote{http://asc.harvard.edu/ciao} tools,
version 4.2. We merged observations from different chips since we are
interested in a morphological analysis and not in the spectral
properties of the merged image. In particular, we are looking for the
features of decreasing emissivity, called `dips', along the M87
jet. We obtained a picture without any systematic offset in the World System Coordinates (WCSs)
of the observations. However, to be sure that the simple merging
procedure was sufficient to guarantee a correct image we checked that
the counts in a circular region on knot F of the merged image was
equal to the sum of the counts of the individual images at the same
position.  We also created exposure maps and merged exposure map of
all the observations used in the analysis to check that the exposure
map was uniform across the regions close to the jet.

\begin{figure}
\centering
\includegraphics[width=1.00\hsize,angle=0,clip]{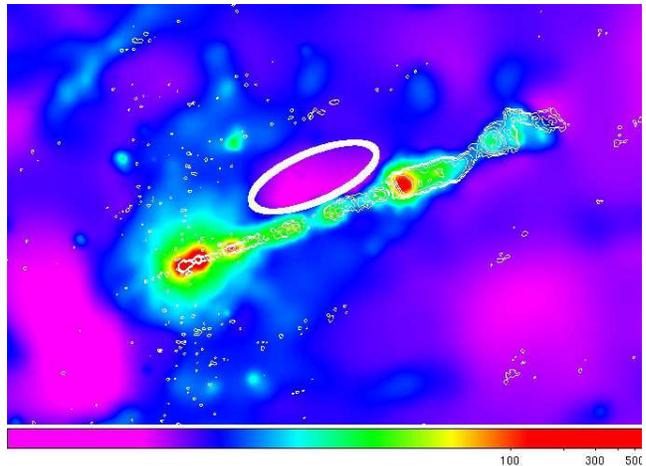}
\caption{An adaptively smoothed image in the $0.3$-$2$~keV energy
  range, with the VLA radio contours superimposed. The lower surface
  brightness region on the north side of the jet is indicated by an
  elliptical region.}
  \label{fig3}
\end{figure}

\subsection{Morphological analysis of the merged data}

In order to perform a detailed analysis of the X-ray surface
brightness near the jet we have split and analysed the merged data
into a few selected energy bands: $0.3-0.8$ keV, $0.8-1.2$ keV,
$1.2-2.0$ keV, $2.0-4.0$ keV, $4.0-7.0$ keV and an additional wider
soft band $0.3-2.0$ keV, to look for an energy dependent
morphology. In Fig.\ref{fig2} we show the merged image in the soft
band $0.3-2$keV, where the most visible dip structures appear; they
are less pronounced at higher photon energies. Close to the jet, the
ISM is cool and dense, producing the complex surface brightness
structures we see in the central region.  The harder X-rays however
arise from hotter and less dense gas in the outer parts of the galaxy,
and the surface brightness above 2 or 3 keV will be produced mostly by
the integrated emissivity along the line of sight rather than by
discrete features near the jet.  Thus we do not expect to see any
indication of a CR cacoon at higher energies.  Between knots F and E
one can note a feature of lower brightness that is visible especially
on the north side of the jet. The analogous dip is present next to
knot F on the south side of the jet, but it is less remarkable. The
mentioned dip structures are more clearly visible for the same data,
both on the north and south side when we apply the adaptive smoothing
tool, {\it csmooth}, as illustrated in Fig. \ref{fig3}. We note here
that the interpretation of the maps is not straightforward due to
confusion with the non-uniform thermal background emission near the
centre of M87.

For Fig. \ref{fig3} we have used csmooth. The csmooth procedure
creates a reference scale map (in this case using the 0.3-2.0 keV
data) and we used this scale map for the other energy
bands \footnote{for the details on how to create maps with csmooth see
  \cite{Balucinska}.}. In the analysis we have also tested other
smoothing techniques provided in the CIAO software: the Quadtree
binning with {\it dmnautilus} and {\it aconvolve} (a simple Gaussian)
in order to compare the resulting pictures, but there was no visible
difference or improvement in comparison to the data treated with {\it
  csmooth}. One should remember here that the adaptive smoothing
procedure in CIAO should not be used for any quantitative analysis as
it can redistribute the flux. When the kernel is variable, the {\it
  csmooth} algorithm has the effect of moving flux from low surface
brightness regions into high surface brightness regions with a
possibility of creating spurious emission, Diehl \& Statler (2006).

\subsection{Hardness ratio maps}

\begin{figure}
\centering
\includegraphics[width=0.70\hsize,angle=-90,clip]{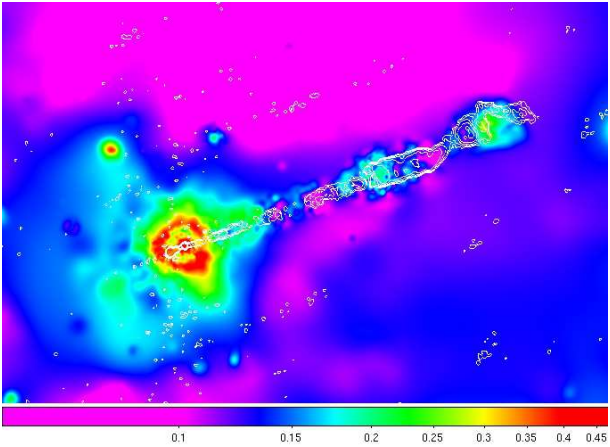}
\caption{The hardness ratio map HR=(2-7 keV)/(0.3-2keV), overlayed
  with the radio contours. The structure at the galaxy centre results
  from the instrumental phenomenon of photon pile up.
\label{fig4}}
\end{figure}

To get additional insight into the weak emission from the dips we
constructd hardness ratio maps of the jet. Let us define the hardness
ratio as

$$HR \equiv {H \over S},$$

\noindent
where $H$ is the hard band counts/pixel and $S$ is the soft band
counts/pixel. The HR maps are derived by dividing the {\it csmoothed}
$H$ and $S$ maps using the CIAO procedure {\it dmimgcalc}. As the
reference soft energy band we have selected the $0.3-2.0$ keV
range. The smoothing pattern for this map was used to smooth the high
energy band $2.0-7.0$ and derive the HR map H($2.0-7.0$)/S($0.3-2.0$),
as presented in Fig. \ref{fig4}. One may see here that the jet has a
harder spectrum than the surrounding gas, and there are softer regions
observed between the jet knots. Also the large dip region from
Fig. \ref{fig2} is apparently harder than the neighbouring galactic
plasma. This is consistent with the harder photons in the dip area
arising from hot gas along the line of sight. No particular spectral
feature which could be ascribed to the jet outer layers or to a CR
cocoon is obvious.

\subsection{The cross-jet intensity profiles}\label{profiles}

\begin{figure}
\centering
\includegraphics[width=1.0\hsize,angle=0,clip]{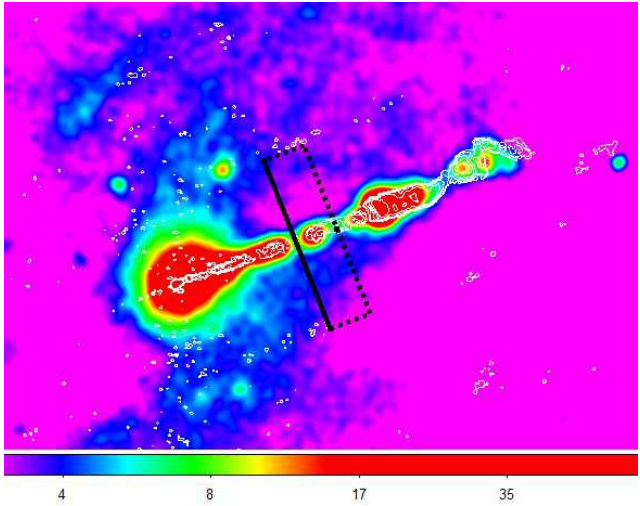}
\caption{The region used for the projection (2.5$^{\prime\prime}$
  width) across knot F in the energy band $0.3-2$ keV. The image is
  binned to a pixel size 0.061" and smoothed with a
  Gaussian of FWHM=1.03$^{\prime\prime}$.
\label{fig5bis}}
\end{figure}

\begin{figure}
\centering
\includegraphics[width=1.0\hsize,angle=0,clip]{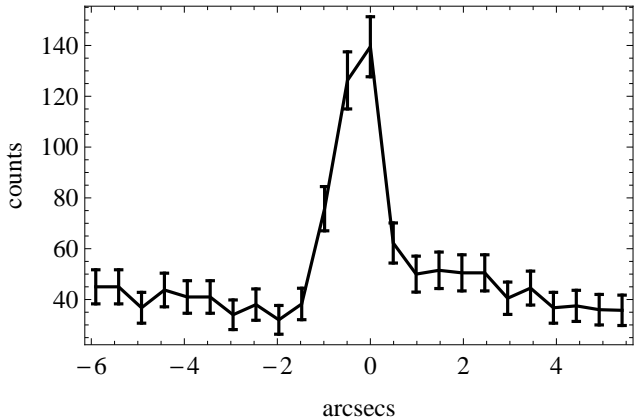}
\caption{The projection across knot F in the energy band $0.3-2$ keV
  for the single longest observation (obsid 5827).
  No binning nor smoothing has been used.
\label{fig5a}}
\end{figure}

\begin{figure}
\centering
\includegraphics[width=0.70\hsize,angle=-90,clip]{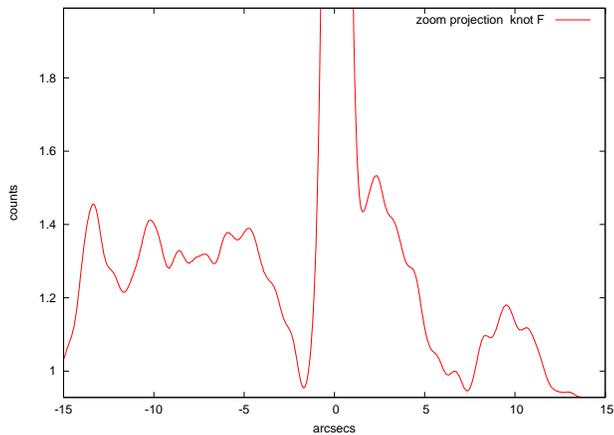}
\caption{The projection across knot F in the energy band $0.3-2$ keV,
  binned with pixel size 0.061" and smoothed with a
  Gaussian of FWHM=1.03$^{\prime\prime}$.
\label{fig5}}
\end{figure}

Using {\it ds9} we have analyzed perpendicular projections across the
jet to visualize the intensity variations.  Additionally, to be sure
that the evaluation of the above mentioned features is not affected by
photon pile up (when two or more photons arrive at the same pixel in
the same frame time and are registered as a single event with the sum
of the photon energies \cite{Davis2001}), or by the spatial
contamination due to flaring activity of HST-1, we have excluded from
the analyzed data the observations from January 2004 up to June
2006. In this period HST-1 - the first jet knot clearly resolved by
Chandra from the nuclear emission - was the site of a huge X-ray flare
in 2005, Harris et al. \shortcite{Harris2009}. To be even more cautious, we
considered projections of the jet starting from knot E, thus avoiding
effects of HST 1 variability. We tested different values of the
projection width in order to optimize the signal to noise while still
being able to resolve the dip. In Fig. \ref{fig5}, we present a
profile perpendicular to the jet located at knot F of all observation and in Fig. \ref{fig5a} of a single observation.  These clearly
exhibits the lower surface brightness near the jet.
 A width of 2.5$^{\prime\prime}$
resulted from these tests (Fig. \ref{fig5bis}.  For this procedure we
have binned the map to a pixel size of 0.061'' and smoothed it with a
1.03$^{\prime\prime}$ FWHM Gaussian.

\begin{figure}
\centering
\includegraphics[width=0.7\hsize,angle=-90,clip]{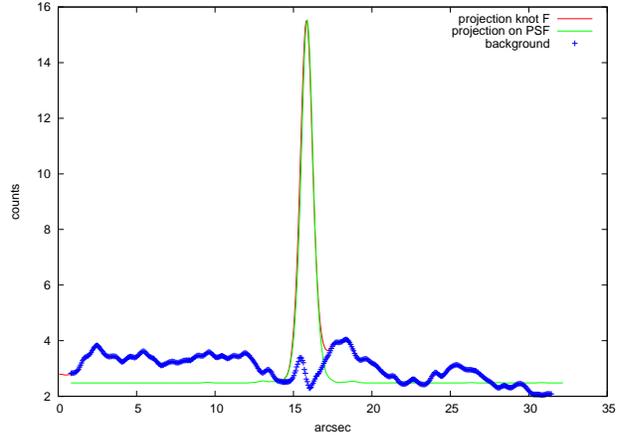}
\caption{The background near the jet (blue curve) is obtained by
  subtracting the modeled PSF projection (green line) from
  the knot F projection (red line).
\label{fig6}}
\end{figure}

As knot F looks like a point source on the map we decided to model it
using the Chandra Point Spread Function (PSF) as a point source
imposed on the local background. To perform the PSF fitting we used
the longest single observation, obsid 5827, in order to extract the
source spectrum that cannot be properly computed if we consider the
merged file. In case of a merged image we don't have a single
aspect solution since it is varying between different observations. We
fitted the spectrum with a simple power law, with a photon index
$\Gamma=2.3$, applying the Galactic absorption modeled as neutral gas
with the Solar abundances and a column density equal $1.94 \times
10^{20} cm^{-2}$, the value determined by the Leiden/Argentine/Bonn
(LAB) Survey of Galactic H I \cite{Kalberla2005}. The PSF map, scaled
to knot F after subtracting the background near the knot, was obtained
running ChaRT and MARX with Ciao 4.2. One should note
that the background level near the knot was fitted here by the
requirement of the best PSF fit to the knot. Then, we derived the PSF
projection and subtracted it from the knot F projection, as presented in
Fig. \ref{fig6}. It is clearly visible that the fitted background
level near the jet is significantly lower than the mean fluctuating
background.  In  Fig. \ref{fig5a} we show a profile of a single
observation which confirms that the observed dip structure is not
affected by the merging procedure of the full data set.

\begin{figure}
\centering
\includegraphics[width=1.0\hsize,angle=0,clip]{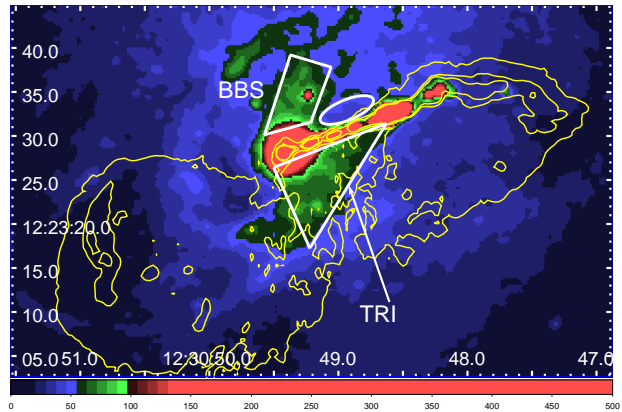}
\caption{The M87 map with 1/4 binning (pix=0.123") and smoothing 0.43"
  FWHM. The yellow 8 GHz radio contours start at 3 mJy/beam and
  increase by a factor of 2 up to 48 mJy/beam.  The white ellipse
  shows the dip position.
\label{fig7}}
\end{figure}

In all our various images there is an asymmetry between the ambient
medium emission north and south of the jet, with a lower surface
brightness to the north. The elongated dips are not visible along the
full length of the jet and between knots E and F are much more
pronounced on the north side of the jet. There is a wealth of
structured emission below 2 keV in the central regions of M87
(e.g. Fig. \ref{fig7}).  The surface brightness dip in the region
along the north side of the jet, in the vicinity of knot F, can be a
manifestation of a cosmic ray cocoon, but it could also be a simple
void in projected background structures. In the next section we
attempt to examine these possibilities.

\section{The soft X-ray structures in the central region of M87}

\subsection{Nature of the features with a decreased brightness}

The central part of M87 is dominated by a bright circular region with
an approximate radius of $50''$ or 3.85 kpc, although there are no
unique sharp boundaries.  Beyond this circle extend the well known `SW
Spur' and the `Eastern Arm' -- for a more extensive description see
Forman et al. \shortcite{Forman2007}. Within the central circular
region a complex brightness distribution is observed, most clearly
seen below 2 keV.  At higher energies, the surface brightness contrast
diminishes. In fact, if we repeat the same analysis in knot F (adopted
for the low energy band and described in sec. \ref{profiles}) for the
$2.0-7.0$ keV band, we recover a decrease in the intensity that is 6
times smaller than the one observed in the band $0.3-2.0$ keV.  With
the exception of the rims around `bubbles' \cite{Forman2007}, it is
difficult to isolate coherent structures on the map, although some
obvious large arcs/filaments can be noted.  Sparks et
al. \shortcite{Sparks2004} have demonstrated a spatial association
between some, but not all, of the X-ray bright features with regions
containing H$\alpha$ emission.  They also suggest that the brighter
X-ray features come from $0.8$ keV gas, a bit cooler than the
$1.4$ keV temperature describing most of the ISM. They suggest that
this cooler gas is associated with the 10$^4$ K gas producing the
observed H$\alpha$ emission.

It is our impression that while there is undoubtedly more than one
structure along some lines of sights, most of the bright regions are
likely to be higher density volumes with a depth roughly equivalent to
their lateral extent.  [If the entire complex of excess brightness
  were to be jet associated, one could imagine a ``fat sausage''
  centered on the jet, with various higher density regions within
  it. We would thus be seeing an almost end-on view of the sausage and
  we would then expect that most lines of sight would pass through
  several regions of higher emissivity \cite{Forman2007}]. Of primary
interest to us are two broad swaths: a roughly triangular region (TRI)
south of the jet and a `broad brush stroke' (BBS) running north of the
nuclear region, loosing its identity after about $10^{\prime\prime}$
(Fig. \ref{fig7}).  The triangular region extends from the south side
of the jet, mostly southerly.  Its top side along the jet runs from
the region around the nucleus out to knot A. We find it curious that
the southern triangle stops abruptly at the jet, and does not continue
north; if it had gone further north, there would be no dip.  It tapers
as it crosses a broad horizontal region well below the jet. When radio
contours are superposed (Fig. \ref{fig7}) it appears that the two
lower sides of the triangle coincide with the edges of both
lobes. While there are no sharp boundaries in the radio brightness, it
is not difficult to pick a contour level that matches the X-ray
excess. In reality, even if there were sharp boundaries of exclusion,
they would most likely be blurred by projection effects: it is highly
likely that there is significant (hidden) radio structures in the 3rd
dimension since it is so difficult to identify continuity between the
inner and outer radio structures.

The northern feature ('BBS') defines the eastern end of the dip and
the jet itself acts as the long southern edge of the dip.  On the
north side of the jet there are no bright radio features, only low
brightness and diffuse emission.  Thus there are no discrete radio
regions which might exclude the hot ISM and explain the existence of
the dip as a normal `radio cavity'.

\subsection{Is the dip associated with the jet?}\label{dipsize}

The low brightness `dip' is identified by the ellipse in
Fig. \ref{fig7}. At 77 pc per arcsec, the elliptical region in the
figure would have a projected size of approximately 400 x 150
pc. Since we are currently assuming the region is related to the jet,
its long axis would be deprojected in the same way as the jet, taken
to be at 15$\pm$5 degrees to the line of sight.  Thus the deprojected
estimate of the size of the region would be 1.5 x 0.15 kpc.

The brightness level in this dip is comparable to the region beyond
the end of the jet and to the area to the south of the outer end of
the jet (i.e.  beyond knot B).  In general, the hardness ratio maps
demonstrate that the spectrum of the dip is similar to the regions
mentioned for the brightness comparisons.  Thus it seems unlikely that
the lower brightness areas could arise from absorption by excess
column density along the line of sight which would be absorbing most
of the emission below 1.5 keV; we would have seen much more extreme
values in the HR maps if that had been the case.

In case the dip is not associated with the jet, we have a
relatively isolated region of 200 to 400 pc in size, with essentially
no excess emission from the denser gas responsible for the higher
brightness areas which dominate the 1 keV maps.

\section{Physical constraints for the jet and CR cocoon}

Let us consider a physical situation with energetic particle
acceleration occurring within the outer layers of the jet. This
process produces CR particles diffusing into the ambient medium
outside the jet, which push back the ISM to form a rarefied CR
cocoon. Below we discuss some constraints for such a process resulting
from the overall energy budget and efficiencies of particle
acceleration, diffusive escape and radiative losses.

\subsection{Energy budget}

Let us compute the energy of the ultrarelativistic particles inside
the region, together with the amount of work required for pushing the
ambient gas (with the adiabatic index $\gamma_{amb} = 5/3$) back to
form the CR cocoon. Assuming that the pressure of this region is equal
to the pressure of the ambient gas, $P=1.3 \times 10^{-9}$ dyn cm
$^{-2}$ \cite{Stawarz2005,Stawarz2006} at the distance of
approximately 650 pc from the galaxy center and the dip volume is
$V=0.15 \times 0.15 \times 1.5$~kpc$^3$, we obtain

\begin{equation}
E = \gamma_{cr} P  \times V \approx 2\ {\times}\ 10^{54}$ ${\rm ergs}  
\label{eq:Equ1}
\end{equation}
\noindent

\noindent
where $\gamma_{cr}=4/3$ is the adiabatic index for a relativistic plasma. $E$ could be somewhat larger than the above estimate if the
cocoon is more extensive but not wholly visible because of excess
X-ray emission along the line of sight. A dynamical time scale for the
formation of the dip is obtained as the ratio between the energy $E$
and the jet's kinetic power, $L_j = \eta \times 10^{44}$~erg/s ($\eta$
is a dimensionless parameter and it grows with the increasing amount
of cold protons in the jet  ($\eta > 1$):

\begin{equation}
t_{dyn} = E/L_j \approx 1.7 \ {\times} \ 10^{10} \, \eta^{-1} \quad {\rm [s]}.
\label{eq: Equ2}
\end{equation}

\noindent

Let us indicate with ${\zeta}$ the fraction of $L_j$ used for the
cocoon generation, i.e. the transfer of jet kinetic energy to
acceleration of cosmic rays and to transport outside the cocoon.
To estimate the maximum efficiency, ${\zeta}_{max}$, of this process
one can compare $t_{dyn}$ with the life time of the jet, $t_j$.  $t_j$
must be greater than the light travel time along the jet (the distance
from the nucleus to knot C , $t_{lt}~\approx$~2kpc).  We thus obtain
${\zeta}_{max} \approx 10^{-1} \eta^{-1}$. This value seems to require
quite a large CR acceleration efficiency for explaining the energetics
of the cocoon. Of course, if the observed jet is a stationary
structure lasting a significant number of $t_{lt}$, $t_j = \rho \times 
t_{lt}$, the required efficiency diminishes to $\zeta =
{\zeta}_{max}/\rho$. With the jet activity time $\sim 10^6$~yrs
evaluated by Bicknell \& Begelman \shortcite{Bicknell96} the required
acceleration efficiency decreases to ${\zeta}_{\rm min} \sim 10^{-4}$,
a value which is quite acceptable from the energy budget point of view.

In order to consider the CR cocoon energy budget in a stationary
condition one can compare the above time scale with the one describing
the diffusive particle escape time scale from the cacoon volume:

\begin{equation}
t_{esc} \sim {V^{2/3} \over \kappa_{eff}} 
\label{eq: Equ4}
\end{equation}

\noindent 
where $V^{2/3}$ is the squared (linear) size of the volume. To obtain
a rough numerical evaluation we assume in Eq.~3 the minimum diffusion
coefficient along the magnetic field (the Bohm diffusion, see
e.g. Drury 1983) for the effective diffusion coefficient,
$\kappa_{eff} \approx \kappa_B \sim r_g(\epsilon) c$. With $r_g =
\epsilon /eB$, one obtains a numerical value $\kappa_B \approx 10^{23}
\epsilon_{GeV} B_{\mu G}^{-1}\, {\rm [cm^2 s^{-1}]}$. Henceforce we
  will use a dimensionless energy $\epsilon_{\rm GeV} \equiv \epsilon
  / (1 GeV)$ and magnetic field $B_{\rm \mu G}$ measured in $\mu {\rm
    G}$.  Then, for a power law cosmic ray spectrum $n(\epsilon)
  \propto \epsilon^{-\alpha}$ within the range $\epsilon_{min} <
  \epsilon < \epsilon_{max}$,

\begin{equation}
t_{esc} \sim 10^{19} \epsilon_{{\rm GeV},i}^{-1} B_{\rm \mu G} \quad {\rm [s]},
\label{eq: Equ5}
\end{equation}

\noindent
where $\epsilon_{i} \approx \epsilon_{max}$ or $\epsilon_{min}$ for
flat ($\alpha < 2$) or steep ($\alpha > 2$) particle spectra,
respectively. One can conclude that for the steep CR spectra with
$\epsilon_{{\rm GeV},i} \le 1$ generated at the jet side boundary
$t_{esc} \ge 10^{19}$~s $>> t_{j} \sim 10^{11}$~s and the {\em
  diffusive escape} process can be neglected if we do not assume
unrealistic conditions in the medium. On the other hand, for flat
particle spectra a substantial difference may appear depending on the
CR main component - electrons or protons. Let us note, following
Stawarz et al. \shortcite{Stawarz2005}, that the characteristic values
of $B$ within the jet can not be much weaker than the equipartition
value of a few $\times 10^{-4}$~G. We also note tht the ambient stellar
light energy density ($\sim 10^{-10}$~erg/cm$^3$) is comparable to the
magnetic field energy density. Thus we take $B_{eff} = 100 \mu$G for
our numerical evaluations within the jet volume.

\subsection{CR electron content}

Following Ostrowski (2000) one can evaluate the electron acceleration
time scale within the boundary layer for highly relativistic
electrons using a well known formula for the acceleration in the
Alfv{\'e}nic turbulence, $t_{acc} = \epsilon^2 (v/c)^2 /
9\kappa_{eff}$. With $v_3 \equiv v/(3000 km/s)$ being a characteristic
MHD turbulence velocity, the acceleration time scale

\begin{equation}
t_{acc,e} \sim 3 \times 10^6 \epsilon_{\rm GeV}  B_{\rm \mu G}^{-1} v_3^{-2} \quad {\rm [s]}
\label{Equ6}
\end{equation}

\noindent
while the radiative loss time scale due to synchrotron and inverse Compton processes can be written as

\begin{equation}
t_{rad,e} \sim  4 \times 10^{17}  \epsilon_{\rm GeV}^{-1} B_{eff,\mu}^{-2} \quad {\rm [s]}
\label{eq: Equ7}
\end{equation}

\noindent
where the ``effective'' magnetic field $B_{eff}^2 \equiv B^2 + 8\pi
U_{amb}$ and $U_{amb}$ is the ambient radiation energy density
measured in the jet rest frame (we assume a Thompson limit for the IC
scattering). Comparing the above two expressions, with $v_3 = 1$ and
$B_{\rm \mu G} = 100$ within the jet volume, yields the maximum
electron energy as $\epsilon_{max} \sim 3.7 \times 10^4$~GeV. Thus the
escape time $t_{esc} \sim 10^{15}$ $B_{\rm \mu G}$~s is much longer
than all other time scales both for the jet volume and for the (weak
$B \sim 1 \mu$G) cocoon. Therefore, independent of the detailed shape
of the particle spectrum, diffusive particle escape can not influence
the cocoon structure if the jet is purely leptonic. Also, it may be
difficult to diffusively transport the accelerated electrons from the
jet into the cocoon, suggesting the likelihood of an energetic proton
content for the cocoon.

One may note that for the medium transported within the jet, the
available acceleration time scale must be smaller than $t_{lt} \approx
2\times 10^{11} \, {\rm [s]} $, the turbulent medium transport time in
the jet. With Eq. \ref{Equ6} this requirement imposes an upper limit
for the accelerated particle energy $\epsilon_{lt} \approx 0.7\times
10^5 B_{\rm \mu G} v_3^2$~GeV.

\subsection{CR proton content}

In eq. \ref{Equ6}, by neglecting the acceleration of the jet velocity
gradient (cf. Ostrowski 2000), we find an upper limit for the proton
acceleration time scale. The radiative loss time scale for protons
(Rachen \& Biermann 1993):

\begin{equation}
t_{\rm rad,p} = 5\times 10^{24} B_{\rm \mu G}^{-2} (1+aX)^{-1} \epsilon_{\rm GeV}^{-1} \quad {\rm [s]},
\label{eq: Equ8}
\end{equation}

\noindent
where $a$ is the ratio of the energy density of the ambient photon
field relative to that of the magnetic field and $X$ is a quantity for
the relative strength of (p,$\gamma$) interactions compared to
synchrotron radiation. Comparing the acceleration and radiative losses
time scales, with $aX = 100$, yields the maximum energy of accelerated
protons $\epsilon_{\rm GeV, max} \approx 1.3 \times 10^7$ ($\sim
\epsilon_{lt}$).  For such energies, or larger in the case of
a contribution to the acceleration process from the neglected
velocity gradient processes or $v_3 >1$, the diffusive time scale
$t_{esc,p} \sim 10^{12}$ s becomes comparable or smaller than
$t_{lt}\sim 10^{11}$ s and one may expect a significant outflow of
cosmic ray protons from the jet into the forming cocoon (note that
evaluation of $t_{esc}$ considered the Bohm diffusion limit leading to
the maximum possible time scale).

A non-negligible density of cold protons, $n_g$, within the jet can
limit the maximum CR proton energy due to (p,p) interactions. With an
approximately constant p-p scattering cross section, $\sigma_{pp}
\approx 3.4\times 10 ^{-26} {\rm cm}^2\, $, the respective scattering
time scale

\begin{equation}
t_{pp} \approx 10^{15} n_g^{-1} \quad {\rm [s]},
\label{eq: Equ9}
\end{equation}

\noindent
is much longer -- for ``reasonable'' densities $n_g \simeq
1$~cm$^{-3}$ -- than the jet travel time $t_{lt}$. Thus, one
does not expect to observe any gamma ray signature of the energetic
protons within the jet, but such emission could be present from the
extended interstellar medium within the M87 galaxy, with a total flux
below $ 0.1 \zeta_{\rm min} L_j \sim 10^{39}$~erg/s, and below the
presently observed gamma ray fluxes from this source.

\section{Final remarks}

In this study we describe an X-ray structure along the M87 jet which
could be an indication of a cosmic ray cocoon. However, any definite
claim is not possible for this source because of the highly
non-uniform X-ray brightness arising from high density structures in
the ISM near the centre of the galaxy. The data show a clear emission
dip along the north side of the jet near knots E and F,  and some weak
indications of decreased emissivity in other parts of the jet. We do
not observe any spectral peculiarity or radio structure related to the
mentioned dip, which would suggest - assuming it is a part of a CR
cocoon - that the pressure inside this structure is provided by the CR
nucleonic component.

Following estimates of Stawarz \shortcite{Stawarz2006} we have
computed the energy of the ultrarelativistic particles inside that
region, assuming that the pressure of this component is equal to the
pressure of the ambient gas. Order of magnitude analysis of the cosmic
ray acceleration and transport processes suggest that the formation of
a cocoon requires the generation of a very energetic proton component,
with a flat energy spectrum extending up to $\sim 10^{16}$~eV. If the
existence of a CR cocoon were to be confirmed, it would imply the
acceleration of protons up to extremely high energies in FRI jets.

A more conventional explanation of the dip might be a large blob of
very hot, rarefied gas which would not contribute significantly to the
soft X-ray emission in this region. It is not clear how to produce
such a large (at least 150 x 400 pc) hot cavity without a significant
amount of molecular gas in the galaxy and without a high star
formation and supernovae rate. Another possibility for explaining the
dip would be if a majority of the X-ray emitting gas is in the form of
clouds or filaments within a the hotter gas. Then the dip could be a
region along a given line of sight free of the denser medium simply by
chance. It would additionally require that the observed dip elongation
along the jet would appear by chance.

However, if we use an explanation such as Scenario 2 for the
structures along the jet we are left with the open question about the
processes acting at the jet side boundary and the interface separating
it from the ambient medium.  With the large tangential velocity jump
across this layer one can not expect that nothing dynamic will happen
there. We may note that there are other objects with regions of
decreased brightness along the jet, like the radio source 3C270 and
its wedge-like depressions in diffuse X-ray surface brightness
surrounding the jets, Worral \shortcite{Worral2010}. Therefore, we
are considering a wider programme of multiwavelength studies in order to
answer questions about physical structures of the jet side boundaries
and of physical processes acting there.

\section{Acknowledgment} 
We thank Edoardo Trussoni for valuable comments, W. Forman and
W. Sparks for discussions on the vagaries of the hot ISM, M. Paolillo,
A. Gibiec and A. Szostek for comments and discussions on data analysis
techniques, and {\L}. Stawarz for valuable criticism and advice. MGD
and MO are grateful for the support from Polish MNiSW through the
grant N N203 380336. MGD is also grateful for the support from Angelo
Della Riccia Foundation. This research is funded in part by NASA
contract NAS8-39073 and NASA grant GOO-11120X.

\end{document}